\documentstyle[prb,aps]{revtex} 
\begin{document}
\draft
\title{First-order rigidity transition on Bethe Lattices}
\author{C.~Moukarzel\footnote{Present address: Instituto de F\'\i sica,
Univ. Fed. Fluminense, 24210-340 Niter\'oi RJ, Brazil. \\
email: {\bf cristian@if.uff.br} }}
\address{ H\"ochstleistungsrechenzentrum, Forschungzentrum J\"ulich,\\
D-52425 J\"ulich, Germany.}
\author{P.~M.~Duxbury}
\address{ Dept. of Physics/Astronomy and Cntr. for Fundamental
Materials Research, Michigan State University, East Lansing, MI
48824-1116.}
\author{P.~L.~Leath}
\address{Dept. of Physics/Astronomy, Rutgers University, Piscataway, NY
08854. }
\date{\today}
\maketitle
\begin{abstract}
Tree models for rigidity percolation are introduced and
solved.  A  probability vector describes the propagation of
rigidity  outward from a rigid border.  All components of this  
``vector order parameter'' are singular at the {\it same} rigidity threshold, $p_c$.
 The infinite-cluster probability
$P_{\infty}$  is usually first-order at $p_c$, but often behaves
as $P_{\infty} \sim \Delta P_{\infty} + (p-p_c)^{1/2}$, indicating critical
fluctuations superimposed on a first order jump.  Our tree models for 
rigidity are in qualitative {\it disagreement} with ``contraint counting'' mean field
theories. In an important sub-class of tree models ``Bootstrap'' percolation and
rigidity percolation are equivalent.

\end{abstract}
\pacs{PACS numbers:61.43Bn, 46.30.Cn, 05.70.Fh} 
\section{ Introduction }
\label{sec:intro}

Soon after the resurgence of interest in percolation phenomena, the elastic constants
of depleted materials were studied.  Although early work suggested\cite{Dege76} that the conductivity
and elasticity exponents were the same, it was soon realised that the elasticity exponents
were usually different\cite{Kawe84} and in particular one must draw a distinction between the elasticity
of systems which have only ``central forces''\cite{Fese84} and those which also have ``bond-bending''
forces.  If a system has bond-bending forces, the percolation geometry is in many
way similar to that of the connectivity percolation problem.  Of interest in this paper is
the fact that when a system is supported by only central forces, the percolation 
geometry is {\it very different} than that occuring in connectivity percolation.  We
illustrate this difference by developing and solving
models for rigidity percolation on trees and by comparing those
models with the analogous results for connectivity percolation on
trees\cite{Fies61}.  Many of the concepts we develop using tree models can be extended
to regular lattices, as will be elaborated upon in the paper. 

There have been several different groups of scientists  and engineers interested in the ability
of central force structures to transmit stress.  Besides its intrinsic interest,
this problem is relevant to the analysis of engineering structures, glasses, granular
materials and gels\cite{Modu95}.  The straightforward way to
study this problem is to construct particular models which have only central forces
and to study the types of structures which support stress.  In the physics community,
the standard model has been lattices composed of Hooke's springs.  Direct solution
of the force equations for these lattices has provided quite variable estimates
of the percolation threshold, and considerable controversy about the 
critical exponents\cite{Datt86,Haro89,HLbu90,Arsa93}.  
In the mathematics community, there has been a long history of attempts to related 
the connectivity of a ``graph'' to its ability to support stress\cite{Lama70,Loye84,Whit88,Hend92}. 
The majority of physicists were unaware, till recently\cite{Modu95,Jath95}, that there
is a rigorous theorem which relates connectivity to rigidity but only
for {\it planar graphs}.  Of more practical importance is the fact that
there is a {\it fast algorithm}\cite{Hend92} by which this theorem can be used to actually
find the infinite cluster\cite{Jath95} and stressed backbone\cite{Modu95} of planar graphs (e.g.
the triangular lattice with central forces).   These results are 
relevant to random lattices\cite{Modu95}, which are in many cases of most practical interest.

There are two different types of mean field theory available
for the rigidity transition.  The first, based on an approximate ``constraint counting'',
predicts a second order transition in the ``number of floppy modes''\cite{Thor83}, and
has been extensively applied to the rigidity of glasses and gels.  However it was realised 
in that paper and recently quantitatively confirmed\cite{Jath95} that the number of floppy modes
per site does
not appoach zero at the percolation point.  There is also a recent continuum
field theory\cite{Obuk95} which predicts first order
rigidity, but the connection between the model parameters and
the lattice parameters relevant to rigidity are not clear in that 
analysis.    The tree models developed here provide a 
more complete mean field theory for the rigidity transition.  We find that
the number of floppy modes is continuous near the rigidity transition, but that
the infinite cluster probability shows a first order jump.  We also find that
one sub-class of our tree models are equivalent to tree models
for bootstrap percolation\cite{Chal79}, although on regular lattices they
are not equivalent.

The paper is arranged as follows.  In the next section (Section II), we introduce
the tree geometry and the vector probability (order parameter) used to
describe the transmission of rigidity from a rigid border.  Section III contains the
detailed analysis of the tree models for both site and bond dilution.  In  Section
IV we discuss the mechanism for first order rigidity and discuss the failings of the
 traditional constraint counting mean field theories in the light of the tree results.
We also calculate
the number of floppy modes, and show that even the second derivative
 is non-singular on trees.  This is not too surprising, since surface bonds
dominate on trees.  Section V contains a brief summary and conclusion.

\section{  The geometry and definition of variables}

The structure of the tree models we consider is illustrated in Fig. 1.
 Following normal convention, we define $z$ to be
the number of branches of the tree (for example in Fig. 1a $z=5$).
 In Fig. 1a, each site of the tree is connected
by only one bond to a neighbouring site.  In general we may have
$b$ bonds connecting neighbouring sites (for example in Fig. 1c, $b=3$).  
 Thus two variables in our analysis are
$z$ and $b$.  A third important variable, $g$, is the number of degrees
of freedom per site and is discussed in the next paragraph.
 The feature of the tree geometry which makes the analysis tractable is that
we can calculate the probability of rigidity along separate branches of 
trees {\it independently}, and then combine the branches of the tree to form the
final Bethe lattice. For example one ``branch'' of the tree of Fig. 1a is
presented in Fig. 1b.  We use the letter $P$, with various subscripts,
for the site probabilities of the entire tree (e.g. Fig. 1a), while
 we use $T$, with various subscripts, to denote the site probabilities
of the branches of the trees.  The qualitative behavior of $T$ and $P$ are
the same, and we concentrate for the most part on the analysis of $T$.

Each node (the sites in Fig. 1a,b and the ellipses
in fig. 1c)  represents a ``joint'' (a point-like node)
 or ``body'' (see below) on a lattice or ``graph'',
 and is assigned a certain number of ``degrees of freedom''.  In connectivity
percolation each node is either connected or disconnected, so it has only one
possible ``degree of freedom''. i.e. if a site is disconnected it has one degree
of freedom, while if it is connected it has no degrees of freedom. If we consider a lattice of
joints connected by central force springs, then each  free joint
 has two translational degrees of freedom in two dimensions and 
three degrees of freedom in three dimensions.  However when we make rigid
clusters, they are rigid ``bodies'' so they also have rotational degrees of freedom.  
For example, a {\it body} in two dimensions has 3 degrees of freedom (two translations
and one rotation), while a body in three dimensions has 6 degrees of freedom (three
translations and three rotations). 
In general, we allow each site to have $g$ degrees of freedom. 
Some practically important values for $g$ are as follows,
\begin{mathletters}
\begin{eqnarray}
g & = & 1               \qquad \qquad \qquad \qquad     \hbox{for
connectivity percolation,} \\
g & = & d               \qquad \qquad \qquad \qquad     \hbox{for a
joint,} \\
g & = & d(d+1)/2        \qquad \qquad   \hbox{  for a body.}
\end{eqnarray}
\end{mathletters}
Here $d$ is the spatial dimension. 
We consider growing clusters from a rigid boundary drawn around the
outer perimeter of the tree.  If a rigid cluster grown from this boundary
continues to grow indefinitely, we are
above the percolation threshold, if it dies out we are below the percolation
threshold.  The behavior on crossing the percolation threshold depends on whether the
transition is first order or second order, as will be discussed further below.
In the case of connectivity percolation, there is only one degree of freedom per node,
and we only have to keep track of the probability that connectivity is transmitted away
from the boundary.  In the case of rigidity
percolation it is necessary to consider a larger set of site probabilities.
In fact, each site may have $0,1,2,...g$ degrees of freedom with respect to the
boundary, so we define the probabilities $P_{0},...P_g$ to be
 the probabilities  that a site 
has between $0$ and  $g$ degrees of freedom (DOF) with
respect to the boundary (a similar definition applies to the branch probabilities $T$). 
{\it For example if $g=3$ } \\
(1)  $P_3 (or\  T_3)$  is the probability that a node has 3 DOF w.r.t.
the border\\
(2)  $P_2 (or\  T_2)$  is the probability that a node has 2 DOF w.r.t.
the border\\
(3)  $P_1 (or\  T_1)$  is the probability that a node has 1 DOF w.r.t.
the border\\
(4)  $P_0 (or\  T_0)$  is the probability that a node has 0 DOF 
w.r.t. the border.

The vectors $P$ and $T$ act as order parameters for the rigidity
percolation problem on trees.  However, it is also possible
to define these quantities on regular lattices and it is likely 
that an algorithm could be developed based on these
probabilities.  In fact for the case of a ``diode
response'', a transfer matrix could be used - this would be a
``directed rigidity percolation'' and might be appropriate for granular
media, where contacts only support compressive forces. 

In many physical problems, it is important to disinguish between a site which is {\it overconstrained} or stressed, $P_B$, and one  which is 
{\it rigid} but not stressed (which has probability $P_D$).
In particular, we have previously defined\cite{Moku95} $P_{\infty} = P_0 = P_D + P_B $ 
to be the ``infinite
rigid cluster'' probability.  This is closely analogous to the
infinite cluster in connectivity percolation\cite{Stau85}.  In this analogy, the overcontrained
or ``stressed'' bonds are analogous to the ``backbone'' in connectivity percolation.
Also, just as the dangling ends in connectivity percolation carry no current, the
dangling ends in rigidity percolation carry no stress. 
However for trees we found it clearer to first concentrate on $P_{infinity}$, so
in this paper we do not discuss $P_B$.

\section{ Diluted Bethe lattices}
Consider Bethe lattices of co-ordination number $z$ as shown in Fig 1.
In general our parameters are $g$ (the number of degrees of freedom per node), $z$
(the co-ordination number - Actually we shall usually use $\alpha =z-1$), $b$ (the
number of bonds connecting each pair of nodes) and $p$ (the probability that
a site or bond is present).  We first do the calculations for
a branch of the trees (see Fig. 1b. for a $b=1$ case) and then join the branches together.  
To illustrate the method we first do the case $b=1$, as illustrated in Figs. 1a,b 
with  site dilution.

\subsection{Site diluted Bethe lattices  with b=1}

On any tree, rigidity can only be transmitted to higher levels of the tree 
if there are enough
bars present to offset the number of degrees of freedom of a newly
added node.  For connectivity percolation only one bar is needed.  If a
node is added to a g=2 tree, two bars are needed to offset the two
degrees of freedom of the added node.  In general, if a node with $g$ degrees of freedom is
added, rigidity is transmitted to the next level of the tree provided
the node is occupied {\it and} provided at least $g$ of the lower level nodes
to which the added node is connected are rigid.  We define the probability
that a node is rigid to be $T_0$. The branch probabilities $T_{k}$ with
$k=0,1...g$ are then given by,
$$ T_{0}=p \sum_{l=g}^{\alpha}{\alpha \choose l} (T_0)^l(1-T_0)^{\alpha-l}$$
$$ T_{1}=p {\alpha \choose g-1} (T_0)^{g-1}(1-T_0)^{\alpha-g+1}$$
\hspace{3in}  ...\\
$$T_{g-1} = p {\alpha \choose 1} T_0 (1-T_0)^{\alpha-1}$$
\begin{equation}
T_g = 1 - \sum_{l=0}^{g-1} T_l
\end{equation}
The left hand side of Eqs. (2) refer to a node at the one higher level than the nodes on
the right hand side.  Since we are
looking for asymptotic probabilities a long way from the rigid boundary,
 we expect the probabilites $T_{l}$ to approach steady state
values upon iteration of Eqs. (2).  
Similar expressions to Eqs.(2) are found when the transition is made from the branch probabilies $T_{l}$ (see Fig. 1c)
to the tree probabilities $P_{l}$ (see Fig. 1a), except that we now combine $z$ branches instead of $z-1$ branches.
Thus we find, for example
\begin{equation}
P_0 = p \sum_{l=g}^z {z \choose l}T_0^l(1-T_0)^{z-l}
\label{eq:eqforPr}
\end{equation}
In fact once we have solved the first of Eq. (2) and have found $T_0$, all of the other components of $P$ and $T$ follow.
In particular, if $T_0$ is first order at a particular $p_c$, then all of the other components of $T$ and $P$ are
first order at the {\it same} $p_c$.  Thus we concentrate on the behavior of $T_0$.  

It is interesting to note that Eq. (3) is {\it the same} as Eq. (2) of \cite{Chal79}  which treats {\it
bootstrap percolation} on trees (with the change of variables $R=1-P$, and $g=m$ and $l=z-m$).  In bootstrap percolation
one considers that ferromagnetic order is propagated {\it only} if each site has at least $m$ ferromagnetic
neighbours.  If we start with a ferromagnetic border, it is clear that Eq. (3), with the above
change of variables, describes the propagation of ferromagnetic order outward from the border.   
The correspondence between bootstrap percolation and rigidity percolation is {\it not exact}
on regular lattices, and it is not clear how to distinguish between these two cases in
a continuum field theory calculation.  

Now we do some detailed solutions to the Eqs. (2).  First we treat some simple solvable cases.  

{\it Connectivity percolation ($g=1$)}\\
In this case the first of Eqs. (2) reduces to that found previously\cite{Fies61}.  For example for $\alpha=3$
\begin{equation}
T_0=p(3T_0(1-T_0)^2 + 3T_0^2(1-T_0) +T_0^3)
\end{equation}
which yields the trivial solution $T_0=0$, and the non-trivial solution
\begin{equation}
T_0= {3 - \surd (4/p-3) \over 2} 
\label{eq:res1forTr}
\end{equation}
The percolation point occurs when the non-trivial solution (\ref{eq:res1forTr})
approaches zero, and this occurs at $p_c=1/3$.  Near $p_c$, $T_0$
approaches zero linearly, so the transition is {\it second order} and the
order parameter exponent $\beta =1$.

In order for the problem to lie in the  ``rigidity percolation'' class, there must
be at least two degrees of freedom per node i.e.  $g \ge 2$.  However when $b=1$, if $\alpha = z-1 =2$,
then $p_c = 1$, as all bonds must be present to order to transmit rigidity.  Thus the simplest non-trivial
case is $g=2$, $\alpha=3$ and $b=1$, which we now treat.\\

{\it Rigidity transition for $g=2$, $\alpha=3$ and $b=1$}\\
>From the first of Eqs. (2), we have,
\begin{equation}
T_0=p(T_0^3 + 3 T_0^2(1-T_0))
\label{eq:res2forTr}
\end{equation}
Of course there is always the trivial solution $T_0=0$.  In addition,
Eq. (\ref{eq:res2forTr}) implies 
\begin{equation}
T_0={3\pm \surd (9-8/p) \over 4}
\end{equation}
To ensure that $T_0=1$ when $p=1$, take the positive root.  The new
feature here is that the square root is negative for $p<p_c = 8/9$, so this root is unphysical below $p=8/9$.
For $p<p_c$, the only remaining real root is $T_0 = 0$, so there is a first order jump in $T_0$ at $p_c = 8/9$.
The magnitude of this jump $\Delta T_0=3/4$.  Note also that on approach to $p_c$ from above,
we find\cite{Chal79}
\begin{equation}
T_0 - 3/4 \sim (p-p_c)^{1/2}
\end{equation}
which illustrates the singular corrections to the first order jump
in $T_0$.  This interesting behavior seems usual for both bootstrap percolation
and for rigidity percolation.  From the second of Eqs. (2), we have,
\begin{equation}
T_1=3pT_0(1-T_0)^2,
\end{equation}
 which has the two solutions, $T_1 = 0$ and the result found by subsituting Eq. (7) for $T_0$ into
Eq. (9). There is thus a first order jump in $T_1$ {\it at the same} $p_c$ as that found for $T_0$.  The size of this jump $\Delta T_1 = 1/8$.  Note that $T_1$ is {\it zero} at $p=1$, so $T_1$ {\it rises} from zero as $p$ is decreases, and peaks at $p=p_c$
Since $T_2 = 1-T_0-T_1$, all components of the vector order parameter are first order, and all
of them have a singular correction near $p_c$ as a consequence of Eq. (8).  

{\it Order of the transition for general $g$, $\alpha$, $b=1$}\\
 In the first of Eqs. (2), there is always the trivial solution $T_0=0$.
After removing that, the following equation holds.
\begin{equation}
1=p\sum_{k=g}^{\alpha} {\alpha \choose k} T_0^{k-1}(1-T_0)^{\alpha-k}
\end{equation}
If $g=1$ (connectivity percolation), there is always a constant term on
the RHS of this equation, and this allows a real solution for arbitrarily small $T_0$, and hence
the transition is second-order.  However, if $g
\ge 2$, the constant term on the RHS is absent and the equation cannot be
satisfied for an arbitrarily small real $T_0$.  Thus there must be a
first order jump in $T_0$ for any $z>g \ge 2$.  It is possible to solve
Eq. (10) to find $p_c$ explicitly in the case $g=\alpha-1$, in which case
the first order jump has magnitude $\Delta T_0 = 1 - 1/(\alpha-1)^2$\cite{Chal79}.  However
in general we resort to numerical methods.
Before describing the numerical results, we  first introduce a matrix method
which allows us to treat general $g,b,\alpha $
 
\subsection{Site diluted Bethe lattices for arbitrary $g,\alpha ,b$}

It is possible to generalise the Bethe lattices described above to cases
where more than one bond connects neighbouring nodes.  In the case of site
dilution, removing a site removes all of the $b$ bonds that enter that
site from a neighbour.  In contrast bond dilution removes one bond
at a time and must be treated differently(see later in this section).  Returning to the site
dilution case, note that if $b \ge g$, rigidity is transmitted
across the tree as soon as connectivity percolation occurs.  This is
because any one connection between two nodes with $b \ge g$ ensures
transmission of  rigidity to the
newly added node, provided of course that the prior node is also rigid
with respect to the boundary.  Thus if $b \ge g$, there are only two possible states
for each node: rigidly connected to boundary and not rigidly connected
to the boundary, and the model is ``trivially'' in the connectivity
percolation class.  In contrast, if there are fewer than $g$ bars
connecting two nodes, more interesting node states are possible, and we
must again consider the full set $ T_0,....T_g$, which allow the possibility
of partial transmission of rigidity.
 We now develop a matrix method to treat the non-trivial cases $1 \le b < g$.

Consider adding a site to a branch of co-ordination $\alpha $.  We label the
sites at the previous level $i=1,...,\alpha $ (for example the lower
ellipse in Fig. 1c would have label $i=1$).  Each of these nodes
may have $l_i=0,1,....,g$ degrees of freedom with respect to the border
(for example the lower ellipse in Fig. 1c has $l_1$ degrees of freedom
with respect to the border).

We start by adding a ``free body'' to the tree, so it has $g$ degrees of freedom with
respect to the boundary.  However, when we add the new higher level body to the tree, 
we also add $\alpha b$ bonds.
But not all of the bonds that are added are ``useful'' in reducing the number
of degrees of freedom of the newly added body with respect to the border.  For example,
if a lower level node already has $g$ degrees of freedom with respect to the border, no matter
how many bonds connect it to the higher level body, it does not produce any constraint of the
newly added body with respect to the boundary.  Therefore we must define the ``number of
useful bonds'', $u$, which lies along any sub-branch.  If a lower level body has zero
degrees of freedom with respect to the border, then every bond is ``useful''.  If the
lower level body has 1 degree of freedom with respect to the border, then the first
bond that is added does not constrain the newly added node, so that only $b-1$ of 
the bonds are useful.  In general if a body has $i$ degrees of freedom,
only $u = b-i$ of the added bonds are useful in producing constraint in the higher level
body.  Thus the probability $Q_u$ that a sub-branch has $u$ useful bonds is given by,
(note that since we are considering $1 \le b <g$, $Q_g = 0$)
\begin{equation}
Q_u = 
\cases{T_{b-u} \qquad \qquad &for $u=1, \cdots ,b$
\cr
 1 - \sum_{v=1}^{b} T_{b-v}  &for $u=0$
} 
\end{equation}
Now each sub-branch adds $u_i$ constraints to the newly added body, so the total number
of constraints on the newly added body is $\sum_{i=1}^{\alpha} u_i$.
Thus the probability that the new node have $k$ degrees of freedom is,
$$T_{0} = p \sum_{l_1=0}^g \sum_{l_2=0}^g .... \sum_{l_{\alpha}=0}^g T_{l_1}T_{l_2}...T_{l_{\alpha}} \theta (g-\sum_{i=1}^{\alpha} u_i)$$
$$T_{k=1,...,g-1} = p \sum_{l_1=0}^g \sum_{l_2=0}^g .... \sum_{l_{\alpha}=0}^g T_{l_1}T_{l_2}...T_{l_{\alpha}} \delta (g-k-\sum_{i=1}^{\alpha} u_i)$$ 
\begin{equation}
T_{g} = 1 - \sum_{l=0}^{g-1} T_l
\end{equation}
Where $\theta $ and $\delta $  are the step function and delta function respectively.

For numerical purposes,  a more convenient way of representing these equations is to
add the $\alpha$ sub-branches one at a time using a matrix method. 
 We define the vector  $\vec{T}^L = (T_0^L,T_1^L,T_2^L,...,T_g^L)$
to denote the probability that the newly added body be in one of its possible constraint ``states'' after the
addition of $L$ sub-branches ($L=1,2...\alpha$).
If we have a free node it has $g$ degrees of freedom so before the addition of any sub-branches, $\vec{T}^0 = (0,0,0...,1)$.
We then have the recurrence relations,
\begin{equation}
T_0^{L+1} = T_0^L(T_0+T_1+..+T_{b}) + T_1^L(T_0+T_1+..+T_{b-1}) + .. + T_{b-1}^LT_0 
\end{equation}
and, for $l=1,2...,g$, 
\begin{equation}
T_l^{L+1} = T_l^L(T_b+T_{b+1}+..+T_g) + T_{l+1}^LT_{b-1} + T_{l+2}^LT_{b-2} + .. + T_{l+b}^LT_0 .
\end{equation}
  Eqs.(13) and (14) may be put into matrix form, so that
\begin{equation}
\vec{T}^{L+1} = \tilde M\vec{T}^L = (\tilde M)^{\alpha }\vec{T}^0
\end{equation}
with\\
$$ 
\tilde M = \pmatrix{
 1      & \beta_1     & \beta_2     & \cdots & \beta_{b}   &   0   & \cdots & 0   \cr
 0      & \alpha      &  T_{b-1}    & \cdots & T_1         &  T_0  & \cdots & 0   \cr
0       & 0           &  \alpha     & T_{b-1} &  \cdots    &       &        &\vdots      \cr
\vdots  & \vdots      & \ddots      & \ddots  & \ddots     &       &        & \vdots     \cr
\vdots  & \vdots      & \vdots      & \ddots  & \ddots     &\ddots       &        &  \vdots    \cr
\vdots  & \vdots      & \vdots      & \vdots  & \vdots     & 0      & \alpha       & T_{b-1}     \cr
0       &    0        &  0          & \cdots &  \cdots     & \cdots & 0     & \alpha	 \cr
} $$\\
where, 
\begin{equation}
\alpha = \sum_{l=b}^g T_l
\end{equation}
and,
\begin{equation}
\beta_k = \sum_{l=0}^{b-k} T_l.
\end{equation}
Finally, we must include the possibility that the site is present or absent, so the 
probability vector obeys,
\begin{equation}
\vec{T} = p (\tilde M)^{\alpha} \vec{T}^0 + (1-p) \vec{T}^0.
\end{equation}
As before, the LHS of Eqs. (18) is the probability vector at the next level of the tree
in terms of the probabilities at the lower levels (which are in the matrix $M$).  

A little algebra shows that Eqs. (18) reproduce the 
$b=1$ equations (Eqs. (2)) as they must.  We illustrate the matrix method with a special case
($b \ne 1$) which is analytically solvable.\\

\noindent {\it A non-trivial solvable case, $\alpha=2$, $g=3$,$b=2$}\\
For $\alpha$, $b=2$, $g=3$ Eqs.(18) yield,\\
$$ 
\pmatrix{T_0 \cr T_1 \cr T_2 \cr T_3 \cr} =
 p \pmatrix{
 1      & T_0+T_1     &  T_0    &  0        \cr
0       & T_2+T_3     &  T_1    & T_0       \cr
0       & 0           & T_2+T_3 & T_1       \cr
0       &    0        &  0      & T_0       \cr}^{2} \pmatrix{0 \cr 0 \cr 0 \cr 1 \cr}  
+  (1-p) \pmatrix{0 \cr 0 \cr 0 \cr 1 \cr}
$$\\
The first two of these equations yield,
\begin{equation}
T_0=p(T_0^2 + 2T_1T_0)
\end{equation}
and
\begin{equation}
T_1=p(2T_0X + T_1^2).
\end{equation}
where $X = T_2+T_3$.  Since the sum of the $T's$ is one, we have $X = 1-T_0-T_1$ and this with Eqs.
(19) and (20) yields,
\begin{equation}
3T_0^2 - 4(2-1/p)T_0 +1/p^2 = 0,
\end{equation}
Solving for $T_0$ yields,
\begin{equation}
T_0= { (4p-2) + 2 \surd ((2p-1)^2 - 3/4) \over 3p}
\end{equation}
Then the argument of the square root becomes negative for $p<p_c$ given
by, $p_c = (1+\surd 3/2)/2 \sim 0.933$, so that  $\Delta T_0=0.619$.

{\it Numerical results for general $b,g,\alpha $}

Results of iterating the matrix Eqs. (18) are presented in Figs. 2-4.  Fig. 2a illustrates
that for $g\le b$, the problem reduces to the connectivity percolation case.  The
transition is second order and  only two components of the vector $T$ ($T_0$ and $T_g$)
are finite.  In contrast, when $b\alpha > g > b$ (see Fig. 2b), all of the components of $T$ can
be finite, although all of them are singular at the same percolation point.   This figure
also illustrates the point that the rigidity transition is first order and we have
selected this case to illustrate the fact that in some case the rigidity transition is
{\it weakly} first order.  

In Fig. 3, we illustrate the dependence of rigidity percolation on the
co-ordination number $\alpha$.  In the case we choose here, $g=2$, $b=1$, the
transition is always strongly first order.  The behavior near $p=1$ is
typical of site dilution on any lattice, because the leading term
in the probability that a site is not rigid with respect to the
boundary, is just the probability that the site is absent, i.e. $1-p$.  As $\alpha$ increases,
the point at which $T_0$ breaks away from $1-p$ tends to $p=0$ as intuitively
expected.

If we start from a rigid border, it is easy to verify that the transmission of
rigidity depends on $\alpha$ and the {\it ratio} $b/g$.  in the limit
$b/g \rightarrow 1$, we have rigidity percolation, while if $b/g \rightarrow 0$,
the transition is at $p=1$ and is completely first order.  Using trees, we
are able to probe various values of $b/g$ and we present results for 
$p_c(\alpha, b/g)$ in Fig. 4.  It is seen that for all cases, $p_c \sim {G(g) \over \alpha}$
for $\alpha \rightarrow \infty$.  We also find that for any $b/g < 1$, the 
transition is first order, and the size of the first order jump increases smoothly
as $b/g$ decreases.

>From the site dilution problem, we conclude that the rigidity transition is
always first order, except in cases where it
trivially reduces to connectivity percolation.  However, there appears
to be a square root singularity superimposed on the first order jump in $T_0$.
  However, on site diluted lattices with $b < g$, the
only rigid clusters are those which are attached to the rigid border.  In contrast in
bond percolation it is possible to have {\it internal} rigid clusters, and the cases
$b>g$ are non-trivial.  Thus we now describe calculations for the transmission
of rigidity in bond diluted trees.

\subsection{ Bond-diluted Bethe lattices (general $g$, $\alpha $ and $b$)}

As for the
site diluted case, we define the vector $\vec{T} = (T_0,T_1,T_2,...,T_g)$.
Now, if there is a total of $b$ possible bonds between two nodes, and if each
is present with probability $p$, then the 
 probability $q(k)$ that $k$ bonds are
actually present is
\begin{equation}
s_k = { b \choose k } p^k (1-p)^{b-k}
\end{equation}
Since the nodes have $g$ degrees of freedom, at most $g$ {\it independent
bonds} can connect two nodes. If $k > g$ bonds connect two nodes, $k-g$
of them will be redundant and the two nodes will form part of
a cluster that is internally rigid. Any number of bonds in excess of $g$ does
not add to the number of independent constraints.  Therefore the
probability $q_k$ that $k$ {\it independent} bonds are present between
two nodes is, for general $g$ and $b$,
\begin{equation}
q_k = 
\cases{ 
s_k \qquad \qquad \qquad &for $k<g$ \cr
\sum_{j=g}^b s_j &for $k=g$        \cr
0 &for $k>g$	
} 
\label{eq:qk}
\end{equation}
As in the site dilution case,  these k bonds are not all ``useful'' in transmitting constraint
from the boundary {\it unless} the sub-branch along which they lie
is at least partially constrained.  In fact if the lower
level node has $i$ degrees of freedom with respect to the boundary,
only $k-i$ of the bonds connecting that node to the newly added 
node actually impose contraint.  Clearly if $k \le i$, the branch
imposes no constraint (with respect to the boundary) on the newly
added node.  We thus define the {\it useful} bonds $u=k-i$ , because they 
are able to propagate contraint outward from the boundary.   
The probability $Q_u$ for a branch to have $u$ useful bonds on it
is then given by,

\begin{equation}
Q_u = 
\cases{ 
\sum_{i=0}^{g-u} T_i q_{i+u} \qquad \qquad &for $u=1, \cdots ,g$
\cr
 1 - \sum_{v=1}^{g} Q_v  &for $u=0$
} 
\end{equation}
Now taking $\alpha $ such sub-branches, the total number $U$ of
useful bars is 
\begin{equation}
U = \sum_{k=1}^{\alpha } u_k
\end{equation}
If $U \ge g$, then the new node body will
be rigid. Otherwise it will have $k=g-U$ degrees of
freedom. Formally we then write
\begin{equation}
T_f = \sum_{u_1=0}^g \sum_{u_2=0}^g \cdots \sum_{u_{z-1}=0}^g
Q_{u_1} Q_{u_2} \cdots Q_{u_{z-1}}  \Phi(f,u_1,u_2,\cdots,u_{z-1})
\label{eq:formally}
\end{equation}
\noindent
 where
\begin{equation}
 \Phi(f,g,z,U) = \cases{ 
 \delta (U - (g-f)) \qquad \qquad &for $0<f\le g$  \cr
\hbox{and}	\cr
\theta (U - g)  \qquad \qquad &for $f=0$  
} 
\end{equation}
Where as in the site case, we have used the step function  and  the kronecker delta to 
ensure that the constraint counting is correct.

As for the site diluted case, we can write the
Eqs. (27) in matrix form.  
\begin{equation}
\vec{T}^{L+1} = \tilde M \vec{T}^{L}
\end{equation}
with\\
$$ 
\tilde M = \pmatrix{
1 & (Q_1+Q_2+\cdots+P_g)     & (Q_2+Q_3+\cdots+Q_g)     & \cdots & Q_g   \cr
0 & Q_0                      &  Q_1                     & \cdots & Q_{g-1} \cr
0 & 0                        &  Q_0                     & \cdots & Q_{g-2} \cr
\vdots  & \vdots             & \vdots                   & \vdots & \vdots \cr
0 & 0                        &  0                       & \cdots & Q_0	 \cr
} $$\\
Again starting from a bare node with $\vec{T}^0 = \{ 0,0,\cdots 0,1\}$, and
after connecting $\alpha $ legs, we get the desired probabilities as
\begin{equation}
\vec{T} = (\tilde M)^{\alpha }  \vec{T}^{0}
\label{eq:nextgen}
\end{equation}
To illustrate the matrix method for the bond case, we again do a solvable case with
$b \ne 1$.

{\it A non-trivial solvable case $\alpha=2, g=3, b=2$}

>From Eqs. (23) and (24), we have,
\begin{equation}
(q_0,q_1,q_2,q_3) = ((1-p)^2,2p(1-p),p^2,0).
\end{equation}
Then from Eq. (25), we have,
\begin{equation}
(Q_0,Q_1,Q_2,Q_3) = (1-(p^2+2p(1-p))T_0-p^2T_1,2p(1-p)T_0+p^2T_1,p^2T_0,0).
\end{equation}
Using these expressions in the matrix equation (29), we have,\\

$$ 
\pmatrix{T_0 \cr T_1 \cr T_2 \cr T_3 \cr} =
  \pmatrix{
 1      & (2p-p^2)T_0+p^2T_1       &  p^2T_0                    &  0                    \cr
0       & 1-(2p-p^2)T_0-p^2T_1     &  2p(1-p)T_0 + p^2T_1       & p^2T_0                  \cr
0       & 0                           & 1-(2p-p^2)T_0-p^2T_1    & 2p(1-p)T_0 + p^2T_1       \cr
0       & 0                           &  0                         &  1-(2p-p^2)T_0-p^2T_1   \cr}^{2} 
\pmatrix{0 \cr 0 \cr 0 \cr 1 \cr}  
$$\\
>From the first of these equations, we find
\begin{equation}
T_0=p^3T_0\{(4-3p)T_0 + 2pT_1\}, 
\end{equation}
while the second implies
\begin{equation}
T_1= 2p^2T_0(1-(2p-p^2)T_0-p^2T_1) + (2p(1-p)T_0+p^2T_1)^2.
\end{equation}
Solving Eqs. (33) and (34) for $T_0$ gives the trivial solution $T_0=0$, and,
\begin{equation}
T_0= {(8p^3-12p+8) + \surd ((8p^3 -12p +8)^2 - 12p^2) \over 6p^3}
\end{equation}
This again becomes imaginary at the rigidity threshold, which we find to
be $p_c = 0.918$, and the first order jump in $T_0$ is, $\Delta T_0 = 0.629$.\\ 

{\it Numerical results for general $b,g,\alpha$}

First we note that the for $b=1$, site dilution and bond
dilution are the same, provided we make the transformation $p_{site} \rightarrow p_{bond}$
and $T_{site} = p_{bond}T_{bond}$, thus we focus attention on $b \ge 2$.

We present numerical results for bond diluted trees in Figs. 5 and 6.  In Figure
5, we show that even when $b>>g$ and many internal rigid clusters can exist
on the trees, the rigidity transition remains first order.  In fact, we have not
found any values of $g$ or $b$ for which the bond diluted trees are second
order, except the trivial case $g=1$.  However the rigidity transition is
weakly first order for $b/g \rightarrow \infty$.  A second interesting feature
of Fig. 5 is the non-monotic behavior of $T_1$.  Nevertheless on all of the trees
we studied, the rigidity transition is unique and first order.   As in Eq. (8),
there appears to be a singular behavior superimposed on the first
order jump in $T_0$.   On the bond
diluted trees, the percolation threshold depends on all
three parameters $g,b,\alpha$, nevertheless there is a simple behavior in the large
$\alpha$ limit (see Fig. 6), so that
$p_c \sim G(g,b)/{\alpha }$ for $\alpha \rightarrow \infty$.  
  
\section{ Mechanism and comparison with other theories}

{\it A mechanism for first order rigidity}

The mechanism for the first order rigidity transition is illustrated in Fig.  7a for an
$\alpha =2$, $g=2$, $b=1$ tree and in Fig. 7b for the bond-diluted triangular lattice.  In these
figures, we have presented a rigid cluster and have indicated a bond which we then remove.  On 
removal of the arrowed bond, both of the rigid clusters ``break'' up into more than 2 rigid subclusters.
In Figure 7a, removal of the arrowed bond leads to $6$ rigid subclusters, while in Fig. 7b, removal of the
arrowed bond leads to the formation of $4$ rigid subclusters.  In both cases we are referring to clusters
of mutually rigid $bonds$.  In contrast, in connectivity percolation, removal of a ``cutting'' or
red bond leads to the break-up of the system into {\it two subclusters}.  On large rigid clusters, 
the removal of a ``cutting'' or red bond usually leads to  formation of many subclusters, and this
``cluster collapse'' provides a mechanism for a first order rigidity transition.  However it does
not {\it ensure} a first order transition, as it depends on {\it how many} clusters are formed
when a cutting bond is removed.  In reverse the phenomenon of cluster collapse is ``cluster-freezing''
in which there is a sudden jump in the average cluster size as many clusters suddenly become
mutually rigid (For example by replacing the arrowed bonds in Fig. 8).  It is likely that these
ideas can be used to develop scaling arguments for the amount of cluster-collapse required for
there to be a first order rigidity transition, and we are currently working in that 
direction.\\

{\it Comparison with constraint counting methods }

 For simplicitiy, consider
first bond percolation for which the argument is simplest. 
 On a {\it regular lattice}, there are N nodes
of co-ordination $z$, with each node having $g$ degrees of freedom and with
$b$ bars connecting each pair of nodes.  Now dilute the bars of the network, with p
the probability that any one bar is present.  Then ``on average'', the number
of degrees of freedom, $fN$, that remain at dilution p is\cite{Thor83},
\begin{equation}
fN = Ng  - pbzN/2 + B,
\end{equation}
where the factor of $1/2$ is due to the fact that each bar is shared between two nodes. $B$ is the
number of bonds that are ``redundant'' in that they are in regions of the lattice which would
be rigid even if they were removed.  The mean field approximation  reduces
to assuming $B=0$, so that $f=g-pbz/2$ and thus $f$ approaches zero at $p_c = 2g/bz$.  
This counting procedure is slightly modified on trees, as the border is rigid
 so every bond which is next to but lower than a node in the tree contributes to the rigidity
of that node (the bonds are not ``shared'' as on a regular lattice).

In this case, the constraint counting is 
\begin{equation}
fN = Ng  - pb\alpha N + B.
\end{equation}
Thus we have the same expression as in Eq. (36), with the replacement $\alpha (tree) \sim z/2 (regular\  lattice)$.  If
we again assume that $B=0$, we find, $p_c(B=0) = g/(b\alpha)$. This estimate is grossly in error when compared with
the actual results for trees (see Fig. 6).  Clearly the stronger the first order transition, the more in error the constraint counting mean field theory becomes.

{\it Global constraint counting }

It has been observed that in two dimensions\cite{Jath95}, although the number of floppy modes
is always continuous, the second derivative of that quantity is singular.  This is based
on counting the number of degrees of freedom {\it in the whole lattice}.  If we do a similar
calculation on trees, the surface bonds dominate, nevertheless it is interesting to
see what the results are.  Thus we have done a calculation which keeps track of the number
of redundant bonds on the trees for all levels going outwards from a rigid boundary.  We have
done the calculation for bond diluted lattices with $b=1$.  In that case, the number of redundant bonds 
$l$ levels away from the boundary is given by,
\begin{equation}
B_l = \alpha^{L-l} \sum_{k=g}^{\alpha} (k-g){\alpha \choose k} (pT_0^{l-1})^k (1-pT_0^{l-1})^{\alpha -k}
\end{equation}
$L$ is the total number of levels in the tree.  The total number of
redundant bonds in the tree is,
\begin{equation}
B = \sum_{l=1}^L B_l.
\end{equation}
>From global constraint counting, we then have,
\begin{equation}
f = g - p\alpha + B/N_s
\end{equation}
$N_s = \alpha^L/(\alpha -1)$ is the number of sites on the $L$ level tree. Results for $f$, $\partial f/\partial p$
and $\partial^2 f \partial p^2$ are presented in Fig. 8.  It is clear from these
calculations that there is no singular behavior in the second derivative of $f$ on trees.  However, there
is a peak in the second derivative, but at a value of $p$ considerabley less than $p_c$.

\section{Conclusions}

We have shown that it is straightforward to develop and analyse
tree models for the transmission of rigidity from a rigid
border.   In order to analyse these models we must, in general,
consider the transmission of ``partial'' rigidity, as partially rigid
structures may lead to rigidity higher up the tree.
Some of the main conclusions of our calculations are\\
1.  Except for some ``trivial'' cases which are equivalent
to connectivity percolation,  the rigidity transition in these systems is {\it first order}.
However there may be a singular piece superimposed upon the first order transition
in the infinite cluster probability, as was explicitly demonstrated in some special cases
(see e.g.  Eq. (8)). \\
2.  Constraint counting mean field theory which ignores redundant bonds is
qualitatively incorrect for trees.   This method does not describe correctly the
nature of the rigidity transition. It can also grossly underestimate
$p_c$, especially if the transition is strongly first order. \\
3.  We have defined a vector order parameter which describes the
number of degrees of freedom two points have with respect to each other.
Although there is the possibility of multiple phase transitions with such
a vector order parameter, we find that there is only one transition on
trees.\\
4.  The number of floppy modes and its first and second derivatives
are non-singular, probably due to the dominance of surface bonds on 
trees.\\
5.  Bootstrap percolation and rigidity percolation are exactly
the same on $b=1$ trees, but different on regular lattices.  It is not
clear, at least to these authors,  to which case (if either), the current
continuum field theory applies\cite{Obuk95}.\\

Taken together with new numerical results in
two and three dimensions\cite{Modu95,Modu96,Jath96}, there is now quite strong evidence that the rigidity
transition on random lattices is often {\it first order}, in contrast to the
large number of earlier papers which have assumed the opposite.  However it
is important to emphasize that the new work using exact constraint counting
is correct for {\it random lattices} while the earlier work was for
regular lattices.  It is still an open question as to whether these
two cases are qualitatively different.\\

\noindent {\bf Acknowledgements}\\
Two of us (CM and PMD) thank the DOE under contract DE-FG02-90ER45418
and the PRF for financial support.  We thank Mike Thorpe for useful discussions
concerning floppy modes.
%
%

%
\centerline{\bf \large Figure Captions}
\figure{\bf 1 \rm The geometry of trees. a) A $z=5$, $b=1$ tree; b) One branch
of the tree of a); c) One branch of a $b=3$ tree.}
\figure{\bf 2 \rm Rigidity percolation of site diluted trees: a) $\alpha=4$,
$g=3$ and $b=3$. The infinite cluster probability and the
probability $T_3$ are plotted.  In this case the behavior is
the same as connectivity percolation, so $p_c = 1/{\alpha}$
and the transition is second order, with $\beta = 1$; b) 
$\alpha=5$, $g=3$ and $b=2$.  $T_0$, $T_1$, $T_2$ and $T_3$
are plotted.  All are first order and singular at the same $p_c$}
\figure{\bf 3 \rm $T_0$ for $g=2$ and $\alpha=1$ for various $\alpha$.  The transition
is always first order}
\figure{\bf 4 \rm $p_c$ as a function of $b/g$ and $\alpha$.  From the
top, the curves are for $b/g = 1/6;1/3;1/2;2/3;1$}
\figure{\bf 5 \rm Rigidity percolation for a {\it bond-diluted} tree
with $\alpha=2$, $g=2$ and $b=40$.  The transition is close to
second order and there is an interesting non-monotonic behavior in
$T_1$}
\figure{\bf 6 \rm $p_c$ for bond-diluted trees.  Curves are for (from the
top) $g=3,b=1$; $g=2,b=1$; $g=6,b=5$; $g=6,b=10$; $g=2,b=10$}
\figure{\bf 7 \rm The effect of removing a bond on the cluster size
distribution.  a) Removing the arrowed bond from this rigid cluster
leads to 6 separate rigid clusters. b) Removing the arrowed bond from
this connected cluster leads to 4 separate rigid clusters.}

\figure{\bf 8 \rm Floppy modes on a bond-diluted tree with $\alpha=6$, $g=3$
and $b=1$.  The number of floppy modes per site is continuous
as are its first ($f'$ and second $f''$derivatives.}

\end{document}